\newif\ifShowKeys
\ShowKeystrue
\ShowKeysfalse

\documentclass[11pt,a4paper]{article} 
\pdfoutput=1
\usepackage[no-natbib-sort]{my-jheppub}

\ifShowKeys \usepackage[notcite]{showkeys} \fi

\usepackage{amsmath, amssymb}

\usepackage{bm,environ,mathrsfs,array,arydshln,booktabs,float,slashed,hyperref}
\usepackage{tensor} 	
\usepackage{wick}
\usepackage[nodayofweek]{date time}

\usepackage{graphicx,epsfig}
\usepackage{epic}

\usepackage[dvipsnames]{xcolor}
\definecolor{maroon}{rgb}{0.8,0.3,0.}

\usepackage{aurical}
\usepackage[T1]{fontenc}

\usepackage{framed}
\definecolor{shadecolor}{RGB}{255, 230, 204}

\allowdisplaybreaks

\newcommand{\be}{\begin{equation}}
\newcommand{\ee}{\end{equation}}
\newcommand{\mc}{\mathcal }
\newcommand{\mb}{\mathbb }

\newcommand{\la}{\label}
\newcommand{\eps}{\varepsilon}

\newcommand{\tr}{\text{Tr} }
\newcommand{\N}{\mathcal{N}}
\newcommand{\imtau}{\text{Im}\,\tau}
\newcommand{\sqcd}{\text{SQCD}}

\title{On the large R-charge $\mc N=2$  chiral correlators and the Toda equation}

\author[a]{Matteo Beccaria} 
\abstract{
We consider $\N=2$ $SU(N)$ SQCD in four dimensions 
and a weak-coupling regime with large R-charge  recently discussed in 
\href{https://arxiv.org/abs/1803.00580v2}{arXiv:1803.00580}. If 
$\varphi$ denotes the adjoint scalar in the $\N=2$ vector multiplet, it has been shown that 
the 2-point functions in the sector of chiral primaries $(\tr\varphi^2)^n$
admit a finite limit when $g_\text{YM}\to 0$ with large R-charge growing like
$\sim 1/g^2_\text{YM}$. The correction with respect to  $\N=4$ correlators is 
a non-trivial function $F(\lambda; N)$ of the fixed coupling $\lambda=n\,g^2_\text{YM}$
and the gauge algebra rank $N$. We show how to exploit the Toda equation 
following from the $tt^*$ equations in order to control the R-charge dependence. 
This allows to determine $F(\lambda; N)$ at order $\mc O(\lambda^{10})$ for generic $N$, 
greatly extending previous results and placing on a firmer ground a conjecture proposed for the 
$SU(2)$ case.  We show that a similar Toda equation, discussed in the past,  may indeed 
be used for the additional sector $(\tr\varphi^2)^n\,\tr\varphi^3$ due to the special mixing properties
of these composite operators on the 4-sphere. We discuss the large R-limit in this second case and 
compute the associated scaling function $F$ at order $\mc O(\lambda^7)$ and generic $N$. Large $N$ factorization
is also illustrated as a check of the computation.
}
\affiliation[a]{Dipartimento di Matematica e Fisica Ennio De Giorgi,\\
Universit\`a del Salento \& INFN, Via Arnesano, 73100 Lecce, Italy} 
\emailAdd{matteo.beccaria@le.infn.it} 

\begin{document}
\maketitle


\section{Introduction}

The search for explicit tests of  AdS/CFT correspondence 
has often induced new applications of old 
non-perturbative theoretical tools aimed at connecting  the two sides of the correspondence.

From the point of view of the dual gauge theory, superconformal invariance implies that natural 
objects to be addressed are 2- and 3-point functions of primary fields. In the early studies of 
type IIB superstring on $\text{AdS}_5\times S_5$ dual to $\N=4$ SYM in 4d, particular attention has  been devoted 
to the protected chiral sector where the supergravity limit is exact due to superconformal
non-renormalization theorems 
\cite{Intriligator:1998ig,DHoker:1998vkc,Lee:1998bxa,DHoker:1999jke,Baggio:2012rr}.
Instead, more general unprotected operators have anomalous dimensions depending non-trivially
on the 't Hooft coupling. Currently, they are efficiently computed by the 
Quantum Spectral Curve
approach \cite{Beisert:2010jr,Gromov:2013pga,Gromov:2017blm}. 

Wilson loops are a (non-local) different class of gauge theory observables which are also 
particularly suited from the point of view of 
testing AdS/CFT. In $\N=4$ SYM, the relevant  observable is  the 
Maldacena-Wilson loop $W(C)$ 
where the six scalar fields are coupled to the loop $C$
in a locally supersymmetric way \cite{Maldacena:1998im}.
At strong coupling, $W(C)$  has a clean dual representation as a disc string amplitude.
In some cases, exact results for $W(C)$
are available using
supersymmetric localization \cite{Pestun:2016zxk}. Indeed, 
it was first conjectured that the $\tfrac{1}{2}$-BPS
circular loop could have been captured by a matrix model 
\cite{Erickson:2000af,Drukker:2000rr}. Pestun proved this claim in \cite{Pestun:2007rz}
by showing that the path-integral computing $W(C)$ in $S^4$
can be obtained from   the contributions at a finite set of critical points. This remarkable feature, due to 
supersymmetry, reduces
the calculation to a finite dimensional (Gaussian) matrix model as first suggested in \cite{Erickson:2000af}.
These exact results give strong coupling predictions that 
open the way to  very non-trivial tests of the AdS/CFT correspondence, as in
\cite{Medina-Rincon:2018wjs}.

If supersymmetry is reduced to $\N=2$, where the low energy effective action is captured by 
Seiberg-Witten theory \cite{Seiberg:1994rs,Seiberg:1994aj}, further exact results are 
available by exploiting once again localization \cite{Teschner:2014oja}. Indeed, 
Pestun's analysis in \cite{Pestun:2007rz}
already includes  the case of pure $\mc N=2$ and mass-deformed $\N=2^*$ SYM on the 4-sphere $S^4$.
With respect to the $\N=4$ theory, there is an additional perturbative 1-loop contribution and  
instanton corrections encoded in Nekrasov partition function
 \cite{Nekrasov:2002qd,Nekrasov:2003rj}. The associated matrix model is no more Gaussian, but nevertheless
 may be treated efficiently. \footnote{
For applications of localization to Wilson loops in $\N=2$ superconformal theories see
\cite{Rey:2010ry,Passerini:2011fe,Russo:2013sba,Fiol:2015spa}.}

Here we are interested in high order perturbative expansion of 
correlators of chiral primaries in $\mc N=2$ superconformal theories in 4d. Although not protected as in $\N=4$, chiral primaries belong to short multiplets and are 
annihilated by all right-chiral supercharges. This implies that, under operator product expansion (OPE), they 
have a (chiral) ring  structure whose dependence on the marginal gauge couplings is non-trivial, and 
encoded in the coefficients appearing in the 2- and 3-point functions. 
In particular, it has been shown how to compute the full set of chiral two-point functions 
in $\N=2$ superconformal $SU(N)$ Yang-Mills theory with $N_f=2\,N$ fundamental hypermultiplets
by  localization \cite{Baggio:2014ioa,Baggio:2015vxa,Gerchkovitz:2016gxx,Billo:2017glv}.

In this context, the authors of  \cite{Bourget:2018obm} have recently 
reconsidered the study of  particular chiral correlators of the form 
\be
\la{1.1}
g_{2n}(\tau,\bar\tau) = 
\langle (\tr\varphi^2)^n\,(\tr\bar\varphi^2)^n\rangle, \qquad \varphi\equiv\varphi(\infty), \ \bar\varphi
\equiv \bar\varphi(0),
\ee
where $\tau=\frac{\theta}{2\pi}+\frac{4\pi\,i}{g^2_\text{YM}}$ is the complexified gauge coupling
and $\varphi$ is the complex adjoint scalar in the $\N=2$ vector multiplet. The 2-point function 
$g_{2n}(\tau, \bar\tau)$ may be computed in the large R-charge limit where \footnote{
The fixed large $n$ coupling $\lambda$ in (\ref{1.2}) follows the notation in \cite{Bourget:2018obm}
and should not be confused with the large $N$ 't Hooft coupling. 
}
\be
\la{1.2}
g\to 0,\qquad \text{with fixed}\ \ \lambda=n\,g^2.
\ee
In the large $n$ limit (\ref{1.2}), it is possible to show that the following ratio is finite
\be
\la{1.3}
F(\lambda; N) = \lim_{n\to \infty}\frac{\left. g_{2n} \right|_\text{SQCD}}
{\left. g_{2n} \right|_{\N=4}}.
\ee
The function $F(\lambda; N)$ has certain special features discussed in \cite{Bourget:2018obm},
like universality and 
exponentiation in the $SU(2)$ case,
and seems an interesting non-trivial object by itself. It has been computed
at order $\mc O(\lambda^3)$ for generic gauge group rank $N$, and at order $\mc O(\lambda^5)$ 
for the specific values $N=2,3,4,5$.  

From a general perspective, it is interesting that non-trivial correlators are found in the zero coupling 
limit, as soon as the R-charge is large $\sim 1/g^2$. Further motivation for the study of 
the limit (\ref{1.2}) is that it falls into the general framework of semiclassical effective field theory  
description of strongly-coupled conformal  theories in sectors of large global charge 
\cite{Loukas:2018zjh,Hellerman:2018xpi}. \footnote{
Additional references about the large charge expansion of general conformal field theories
and in AdS/CFT context may be found in \cite{Hellerman:2015nra,Alvarez-Gaume:2016vff,Monin:2016jmo,
Hellerman:2017sur,Loukas:2016ckj,Hellerman:2017efx,Banerjee:2017fcx,
Loukas:2017lof,Hellerman:2017veg,Jafferis:2017zna,Loukas:2017hic,Tseytlin:2003ii}. 
}

As shown in \cite{Baggio:2014ioa}, 
the correlator (\ref{1.1}) obeys a Toda equation equivalent to the 4d version of the 
two-dimensional topological-anti-topological fusion $tt^*$ equations \cite{Cecotti:1991me,Cecotti:1991vb}.
In the analysis of \cite{Bourget:2018obm}, this fact
is used as check of the localization computations. In this paper, we consider the higher order calculation of the 
function $F(\lambda; N)$ for generic $N$. To this aim, we shall exploit the Toda equation as a 
strong constraint to 
control the exact R-charge dependence order by order in the weak coupling expansion. This approach will 
turn out to be quite effective for the calculation of the ratio (\ref{1.3})  and  its large R-charge 
limit. In particular,  we shall present the expression of $F(\lambda; N)$ at order $\mc O(\lambda^{10})$, valid
for generic $N$. As a byproduct, our results support the exponentiation property observed for the $SU(2)$
theory in \cite{Bourget:2018obm}.

A further important issue concerns which type of correlators may generalize those in  (\ref{1.1}),
while still admitting a 
sensible large R-charge limit in the regime (\ref{1.2}). As we explained, 
it is quite important to control the R-charge dependence by a (possibly modified) Toda equation. 
For $SU(N)$ theories this simple decoupling of the $tt^*$ equations has been discussed in 
\cite{Baggio:2015vxa,Gerchkovitz:2016gxx} for the correlators 
\be
\la{1.4}
g_{2n}^\Phi(\tau,\bar\tau) = 
\langle (\tr\varphi^2)^n\,\Phi\,(\tr\bar\varphi^2)^n\,\bar \Phi\rangle,
\ee
and expected to be valid for suitable chiral primaries $\Phi$ at 2-loops, but violated at 
3-loops and higher order in general \cite{Gerchkovitz:2016gxx}. 
This is related to the  
mixing pattern that the conformal anomaly induces when chiral primaries are mapped from $\mb R^4$
to the 4-sphere $S^4$ -- where
localization is naturally worked out. Here, we show that the case $\Phi = \tr\varphi^3$ is special by 
exploiting the normal ordering properties of the composites $(\tr\varphi^2)^n\,\tr\varphi^3$
and their relation with mixing \cite{Billo:2017glv}.
We prove by explicit localization computations that the correlators in (\ref{1.4}) obey in this case
a modified decoupled Toda equation, first suggested in \cite{Baggio:2015vxa}. Besides, we exploit it 
to compute at order $\mc O(\lambda^7)$ and generic $N$ 
the function $F^\Phi$ expressing a ratio analogous to (\ref{1.3}).
Large $N$ factorization
is also illustrated as a check of the computation.

It would be interesting to extend these calculations to other gauge algebras and to 
more complicated correlators  like the one-point functions of chiral primaries in presence of Wilson loops,
recently treated by localization in 
\cite{Billo:2018oog}. Besides, it would be desirable to understand what is the meaning of the proposed
large R-charge limit from the AdS/CFT perspective, {\em i.e.} whether it has a useful 
dual gravity interpretation.

The detailed  plan of the paper is the following. In Sec.~\ref{sec:chiral} we briefly  recall
the main definitions, previous results on chiral 2-point functions, and their large R-charge limit.
In Sec.~\ref{sec:loc} we introduce the basic tools needed to compute chiral correlators by localization, the map 
to the matrix model observables, and also give some explicit examples at 5-loop order.  
In Sec.~\ref{4} we explain how to use the Toda equation to control the R-charge dependence
in the $(\tr\varphi^2)^n$ sector and derive our main result, {\em i.e.} the function $F(\lambda; N)$ at 
order $\mc O(\lambda^{10})$ and generic $N$. Finally, in Sec.~\ref{5} we 
discuss the sector $(\tr\varphi^2)^n\,\tr\varphi^3$, illustrate a modified Toda equation and test it 
against low level explicit correlators, and obtain the associated large R-charge scaling function 
at order $\mc O(\lambda^{7})$ and generic $N$. Sec.~\ref{sec:largeN} is devoted to a brief
discussion of the large $N$ expansion of our results. Various appendices collect technical details
and further discussions.

\section{Chiral correlators in $\N=2$ superconformal theories}
\la{sec:chiral}

Four dimensional superconformal $\N=2$ theories admit chiral primary operators $\Phi_I$
which are primary fields annihilated by half of the supersymmetry charges, $[\overline Q, \Phi_I]=0$.
In terms of the R-symmetry group $SU(2)_R\times U(1)_r$, and dilatation eigenvalue $\Delta$,
they have vanishing $SU(2)_R$ isospin and abelian R-charge $r$ 
saturating the unitarity bound $\Delta=\frac{r}{2}$. \footnote{Chiral primaries are usually assumed to be
scalar fields with Lorentz spins $(j_1,j_2)=(0,0)$. Actually, one of the two Lorentz spin is automatically
zero, while the other is zero in Lagrangian theories and expected to be zero in general, see 
\cite{Buican:2014qla}. In general, they can be detected from the superconformal index, see for instance
the review paper \cite{Rastelli:2016tbz}.
}
The 2-point function between chiral primaries is 
\be
\la{2.1}
\langle\Phi_I(x)\,\bar\Phi_{\bar J}(0)\rangle = \frac{g_{I\bar J}}{|x|^{2\Delta}},
\ee
where the coefficient $g_{I\bar J}$ depends on the marginal couplings. The operator product expansion
is non singular and reads
\cite{Lerche:1989uy}
\be
\la{2.2}
\Phi_I(x)\,\Phi_J(0) = C\indices{^K_{IJ}}\,\Phi_K(0)+\dots,
\ee
where the chiral ring coefficients $C\indices{^K_{IJ}}$ enter the 3-point functions, as usual. If 
there are exactly marginal couplings, the space of marginal deformations has complex coordinates 
$(z_i, \bar z_i)$ and the infinitesimal change of the action $\delta S = \delta z\,\int d^4 x\,\mc O_i(x)$
(plus a similar antiholomorphic part) preserves $\N=2$ supersymmetry if $\mc O_i$ are dimension 4 
descendants of chiral primaries $\varphi_i$ with $r=4$. Their Zamolodchikov metric $g_{i\bar\jmath}$
appearing in (\ref{2.1}) is K\"ahler with a potential given by the logarithm of the partition function 
on $S^4$ regulated in order to preserve the massive supersymmetry algebra $OSp(2|4)$
\cite{Gerchkovitz:2014gta}
\be
\la{2.3}
g_{i\bar\jmath} = \partial_i\,\partial_{\bar \jmath}\,\log Z_{S^4}.
\ee
If the partition function $Z_{S^4}$ may be computed by localization, this provides the Zamolodchikov metric
and the 2-point function of the $\Delta = 2$ chiral primaries. 

This approach has been exploited in \cite{Baggio:2014ioa,Baggio:2014sna} discussing mainly 
the example of $SU(2)$ SQCD, {\em i.e.} YM with 4 fundamental hypermultiplets. In that 
case, the  chiral ring has one primary $\varphi_{2n}\sim (\tr \varphi^2)^n$ for each dimension
$\Delta = 2\,n$,
where $\varphi$ is the complex adjoint scalar in the $\N=2$ vector multiplet. Formula  (\ref{2.3}) 
may be fully {\em extended} to compute the 2-point function $g_{2n}(\tau, \bar\tau)$ in 
\be
\la{2.4}
\langle\varphi_{2n}(x)\,\overline\varphi_{2n}(0)\rangle = \frac{g_{2n}(\tau, \bar\tau)}{|x|^{4n}},
\ee
where $\tau$ is the Yang-Mills complexified coupling 
\be
\la{2.5}
\tau = \frac{\theta}{2\,\pi}+\frac{4\,\pi\,i}{g^2}.
\ee
To this aim, the $n=1$ input, taken from (\ref{2.3}), is fed into the following 
(semi-infinite) Toda equation \footnote{To be precise,
(\ref{2.6}) 
may be mapped to the Toda equation by a simple change of variable.}
\be
\la{2.6}
\partial_\tau\partial_{\bar\tau}\log g_{2n} = \frac{g_{2n+2}}{g_{2n}}-\frac{g_{2n}}{g_{2n-2}}-g_2.
\ee
This equation is a special case of the 4d analogue of the two-dimensional 
$tt^*$ topological fusion equations \cite{Cecotti:1991me,Cecotti:1991vb}  
for the $\N=(2,2)$ chiral ring and derived in 4d using
superconformal Ward identities in \cite{Papadodimas:2009eu}. \footnote{Notice that the 2-point functions
$g_{2n}$ are enough to compute the {\em extremal} 3-point function 
$\langle\varphi_{2m}(x_1)\varphi_{2n}(x_2)\overline\varphi_{2m+2n}(y)\rangle$ 
due to the associativity properties of the chiral ring algebra. 
} 

These results have been extended in \cite{Baggio:2015vxa} to $SU(N)$ SQCD, with $N_f=2N$
massless fundamental hypermultiplets. The extension is not trivial because, 
altough there is again only one marginal coupling, the structure of the chiral ring is more complicated.
In the $SU(N)$ theory, a generic chiral primary takes the form 
\be
\la{2.7}
\varphi_{\{n_1, \dots, n_{N-1}\}} \sim \prod_{\ell=1}^{N-1}\left(\tr\varphi^{\ell+1}\right)^{n_\ell}.
\ee
The $tt^*$ equations are now an infinite set of coupled, non-linear  equations for matrix-valued
objects whose dimension increases with the conformal dimension. Due to the special nature of 
$\varphi_2=\tr\,\varphi^2$,
which is the only one with $\Delta=2$,  it has been proposed that the $tt^*$
equations may be decoupled into separate Toda equations \cite{Baggio:2015vxa}. 
In particular this has been conjectured to happen for special 
primaries $\Phi$ -- called $C_2$-primaries -- connstructed such that the two point functions
\be
\la{2.8}
g_{2n}^\Phi = \langle \Phi^{(n)}(1)\,\bar \Phi^{(n)}(0)\rangle,\qquad \Phi^{(n)} = \left(\tr\varphi^2\right)^n\,\Phi,
\ee
obey the modified Toda equation, cf. (\ref{2.6}) , 
\be
\la{2.9}
\partial_\tau\partial_{\bar\tau}\log g_{2n}^\Phi = \frac{g_{2n+2}^\Phi}{g_{2n}^\Phi}
-\frac{g_{2n}^\Phi}{g_{2n-2}^\Phi}-g_2.
\ee
Relation (\ref{2.9})  was tested at 2 loops \footnote{To be clear, this means the correction $\mc O(g_\text{YM}^4)$
with respect to the tree level value.} 
in $SU(3)$ and $SU(4)$ theories by using conventional Feynman
diagrams, see for instance \cite{Penati:1999fr,Penati:1999ba,Penati:2000zv}.
In order to push the test to higher order, in \cite{Gerchkovitz:2016gxx}
is has been shown how to use localization for the computation of rather general chiral 2-point functions
(and related extremal 3-point functions). \footnote{See also \cite{Fucito:2015ofa} 
for a discussion in terms of Alday-Gaiotto-Tachikawa duality.
}
For each chiral primary $\varphi_i$
(marginal or not) it is possible to deform the partition function on $S^4$ by a term associated with the 
complex coupling $\tau_i$ such that 
\be
\la{2.10}
\langle\varphi_i(\text{N})\,\bar\varphi_{\bar\jmath}(\text{S})\rangle_{S^4} = \partial_{\tau_i}
\partial_{\bar\tau_{j}}\log Z_{S^4}(\bm{\tau}, \bar{\bm{\tau}}),
\ee
where N,S are the north and south poles of $S^4$. Since $\N=2$ supersymmetry is preserved, it is possible to 
compute $\log Z_{S^4}(\bm{\tau}, \bar{\bm{\tau}})$ by minor modifications of the usual 
localization formulas at weak-coupling, especially when instanton corrections are not a concern
as in our weak-coupling expansions.

A well known complication is that we are interested in flat space correlators, while  (\ref{2.10})
 is computed on the 4-sphere and, due to conformal anomaly,  there are important differences.
This is expressed concisely in terms of a mixing 
\be
\la{2.11}
\varphi_{\mb R^4}\to \varphi_{S^4}^{\Delta}+c_1\,R\,\varphi_{S^4}^{\Delta-2}
+c_2\,R^2\,\varphi_{S^4}^{\Delta-4}+\dots,
\ee
where $R$ is the Ricci scalar. The explicit mixing coefficients may be found by a Gram-Schmidt
orthogonalization of the operators on $S^4$.
 \footnote{In general, multiple trace operators are mixed with single trace
operators, at finite rank $N$.
A special solution at large $N$ is discussed in \cite{Rodriguez-Gomez:2016cem} and applied to check 
the chiral primary correlator with a Wilson loop computed in \cite{Giombi:2009ds}
with a 2-matrix model proposal. The large $N$ limit of correlators is also studied in 
\cite{Baggio:2016skg,Rodriguez-Gomez:2016ijh,Pini:2017ouj}.
} In general, this procedure leads to far from explicit results. One important exception is the case of 
$g_{2n}$ in (\ref{2.4}). In this case, in \cite{Gerchkovitz:2016gxx} it is shown that ($Z\equiv Z_{S^4}$)
\be
\la{2.12}
g_{2n}(\tau, \bar\tau) = \frac{1}{Z}\,\frac{\det M^{(n)}}
{\det M^{(n-1)}},
\qquad M^{(p)}_{ab} = \partial^a_\tau\partial^b_{\bar\tau} Z, \ \ a,b=0,\dots, p.
\ee
For our purposes, an important remark is that (\ref{2.12}) implies the Toda equation (\ref{2.6}) 
\cite{Gerchkovitz:2016gxx}. Explicit 3 loop calculations carried on in $SU(3)$ and $SU(4)$
theories in \cite{Gerchkovitz:2016gxx} suggests that the generalized relation (\ref{2.9}) 
is not valid at all orders for the $C_2$ primaries introduced in \cite{Baggio:2015vxa}, {\em i.e.}
the $tt^*$ equations do not decouple. \footnote{To clarify what happens, let us recall that the construction
in \cite{Baggio:2015vxa} is based on the chiral primary tower $(\tr\varphi^2)^n\,\mc O_I$, where $\{\mc O_I\}$
is a basis for fixed dimension chiral primaries with definite orthogonality properties. Its construction requires
a rotation in field space that happens to be coupling-dependent starting at three loops. This extra $\tau, \bar\tau$
dependence spoils the decoupling property. An alternative point of view will be illustrated in 
Section~\ref{5}, see also App.~(\ref{app:dec}).
}

\subsection{Large R-charge limit of $\N=2$ correlators}

As we mentioned in the Introduction,  the authors of \cite{Bourget:2018obm} have recently
proposed to study the chiral correlators $g_{2n}$ in (\ref{2.4})  by considering the 
zero gauge coupling limit $g\to 0$ while simultaneously increasing 
large R-charge $\sim 1/g^2$, see (\ref{1.2})  Working in $SU(N)$ SQCD, they proved that 
in this limit one has 
\be
\la{2.14}
g_{2n} = F(\lambda; N)\,\left(\frac{\lambda}{2\,\pi\,e}\right)^{2n}\,n^{\frac{N^2-1}{2}}+\dots,
\ee
where $F(\lambda; N)$ is the asymptotic ratio to the $\N=4$ correlators defined in (\ref{1.3}).
In the following discussion, it will be convenient to define the  fixed R-charge ratios
\be
\la{2.15}
F(g, n; N) =\frac{\left. g_{2n} \right|_\text{SQCD}}
{\left. g_{2n} \right|_{\N=4}},\qquad 
F(\lambda; N) = \lim_{n\to \infty} F(\sqrt{\lambda/n}, n; N).
\ee
The perturbative expansion of $F(g, n; N)$ is in integer inverse powers of 
\be
\la{2.16}
\imtau = \frac{4\,\pi}{g^2}.
\ee
One of the results of \cite{Bourget:2018obm} are the  explicit first two corrections
\begin{align}
\la{2.17}
F(g, n; N)  &= 
1-\frac{9\,n\,(N^2+2\,n-1)}{4\pi^2\,(\imtau)^2}\,
\zeta(3) \notag \\
&+ \frac{5\,n\,(2\,N^2-1)\,(3\,N^4+(15\,n-3)\,N^2+20\,n^2-15\,n+4}{
4\pi^3\,N\,(N^2+3)\,(\imtau)^3}\,\zeta(5)+\dots.
\end{align}
From this expression, we obtain in the limit (\ref{1.2}), see (\ref{2.15}),
\be
\la{2.18}
F(\lambda; N) = 1-\frac{9\,\zeta(3)}{32\,\pi^4}\,\lambda^2+
\frac{25\,(2\,N^2-1)\,\zeta(5)}{64\,\pi^6\,N\,(N^2+3)}\,\lambda^3+\dots.
\ee
Explicit results at order $\mc O(\lambda^5)$ are presented in \cite{Bourget:2018obm} for the 
specialized cases $N=2,3,4,5$. 
At higher orders, the function $F$ contains products of $\zeta$-functions. 
Remarkably, for the $SU(2)$ theory, one can write
\be
\la{2.19}
F(\lambda; 2) = \exp\bigg(
-\frac{9\,\zeta(3)}{32\,\pi^4}\,\lambda^2+\frac{25\,\zeta(5)}{128\pi^6}\,\lambda^3-\frac{2205\,\zeta(7)}
{16384\,\pi^8}\,\lambda^4+\frac{3213\,\zeta(9)}{32768\,\pi^{10}}\,\lambda^5+\dots
\bigg),
\ee
with only simple $\zeta$-functions in the exponent. This feature has been conjectured to hold at all
orders in the small $\lambda$ expansion. For $N>2$, only terms involving $\zeta(3)$ are supposed to 
exponentiate, {\em i.e.} all dependence on the transcendental $\zeta(3)$ is captured by the first term
in (\ref{2.19}) and is independent on $N$, as follows from (\ref{2.17}).

\section{Localization computation of the chiral correlators}
\la{sec:loc}

In this Section, we briefly recall the available tools to evaluate chiral correlators by localization, the map 
to the matrix model observables, and also give some explicit examples for later purposes.

\subsection{Matrix model $\N=2$ partition function}

Let us consider the $\N=2$ $SU(N)$ Yang-Mills theory with $N_f=2N$ fundamental hypermultiplets (SQCD).
The celebrated partition function resulting from localization  reads \cite{Pestun:2016zxk}
\be
\la{3.1}
Z_\sqcd = \int d^N\,a\,\delta(\sum_n a_n)\,
\Delta(a)\,e^{-2\,\pi\,\imtau\,\sum a_i^2}\,|Z_{1-\text{loop}}|^2\,|Z_\text{inst}|^2,
\ee
where $a_n$ are the eigenvalues of the traceless $N\times N$ matrix $a$ related to the vacuum expectation value
of the adjoint scalar $\varphi$ in the $\N=2$ vector multiplet. The function 
$\Delta(a) = \prod_{n<m}(a_n-a_m)^2$ is the Vandermonde determinant. The 1-loop contribution to the partition function is 
\be
\la{3.2}
|Z_{1-\text{loop}}|^2 = \frac{\prod_{n<m}H(a_n-a_m)^2}
{\prod_n H(a_n)^{2N}},\qquad H(x) = \prod_{n=1}^\infty
\left(1+\tfrac{x^2}{n^2}\right)^{n^2}\,e^{-\frac{x^2}{n}}.
\ee
Finally, $Z_\text{inst}$ is the Nekrasov partition function   \cite{Nekrasov:2002qd,Nekrasov:2003rj}
evaluated with equivariant $\Omega$-deformation $\eps_{1,2}$ parameters
equal to the inverse radius of $S^4$. In the limit (\ref{1.2}) this contribution may be neglected and we shall
not discuss it further.

\subsection{The $\N=4$ case and Gaussian correlators}

The partition function of $\N=4$ SYM is obtained by  removing the 1-loop and instanton terms
and evaluates to the simple expression
\be
\la{3.3}
Z_{\N=4} = \frac{(2\pi)^\frac{N-1}{2}}{\sqrt{N}}\,\frac{G(N+2)}{(4\,\pi\,\imtau)^\frac{N^2-1}{2}},
\ee
where $G(N)$ is the Barnes G-function. \footnote{We remind that for integer $N$ one has 
$G(N+2) = \prod_{n=1}^N n!$.} For our applications, closely following the neat
analysis in \cite{Billo:2017glv,Billo:2018oog},
 it is convenient to rescale the matrix $a$ 
in order to put the classical part of the partition function in standard Gaussian form $e^{-\tr a^2}$.
This requires
\be
\la{3.4}
a\to \left(\frac{g^2}{8\,\pi^2}\right)^{1/2}\,a.
\ee
Going to the basis $a = \sum_{\ell=1}^{N^{2}-1}a^{\ell}\,t^{\ell}$, with 
$\tr t^{a}=0$ and $\tr (t^{a}\,t^{b}) = \tfrac{1}{2}\,\delta^{ab}$, 
one can show that the relevant measure is flat 
$\prod_{n=1}^{N}da_{n}\,\Delta(a) = \text{c}_{N}\,\prod_{\ell=1}^{N^{2}-1}da^{\ell}$. Thus,
matrix model expectation values are computed as Gaussian averages 
$\langle F(a)\rangle = \int da\,e^{-\tr a^{2}}\,F(a)$ where the normalization
$da = \prod_{\ell=1}^{N^{2}-1}\frac{da^{\ell}}{\sqrt{2\pi}}$ is chosen in order to have
$\langle 1 \rangle=1$.
Explicit calculations are performed by first applying Wick 
contractions to pair all $a$'s. Then, contractions are computed by 
using the Gaussian matrix model relation $\langle a^{p}\,a^{q}\rangle = \delta^{pq}$. This makes 
a pair of $t^{\ell}$ appear and reduction is achieved by  using  standard $SU(N)$ identitities 
\cite{Billo:2017glv}.
With the definition 
\be
\la{3.5}
t_{n_1, n_2, \dots} = \langle \tr (a^{n_1})\,\tr (a^{n_2})\dots\rangle, 
\ee
one finds up to level $\sum_i n_i = 8$ the expressions   
{\small
\begin{align}
\la{3.6}
t_2 &= \frac{N^2-1}{2}, & t_{2,2} &= \frac{N^4-1}{4}\notag \\
t_4 &= \frac{(N^2-1)\,(2\,N^2-3)}{4\,N}, & t_6 &= \frac{5\,(N^2-1)\,(N^4-3\,N^2+3)}{8\,N^2}, \notag \\
t_{3,3} &= \frac{3\,(N^2-1)\,(N^2-4)}{8\,N}, & t_{4,2} &= \frac{(N^2-1)\,(N^2+3)\,(2\,N^2-3)}{8\,N}, \notag\\
t_{2,2,2} &= \frac{(N^4-1)\,(N^2+3)}{8}, & 
t_8 &=  \frac{7 (N^2-1) (2 N^6-8 N^4+15 N^2-15)}{16 N^3}, \notag \\
t_{2,6} &= \frac{5 (N^2-1) (N^2+5) (N^4-3 N^2+3)}{16 N^2}, & t_{3,5} &= \frac{15 \
(N^2-4) (N^2-2) (N^2-1)}{16 N^2}, \notag \\
t_{4,4} &= \frac{(N^2-1) (4 \
N^6+20 N^4-99 N^2+135)}{16 N^2}, & t_{2,2,4}&= \frac{(N^2-1) (N^2+3) \
(N^2+5) (2 N^2-3)}{16 N}, \notag \\
t_{2,3,3}&= \frac{3 (N^2-4)(N^2-1) \
(N^2+5)}{16 N}, & t_{2,2,2,2}&= \frac{(N^4-1)(N^2+3)(N^2+5)}{16}.
\end{align}
}

\subsection{$\N=2$ as a perturbation around $\N=4$ SYM}

As is well known, see for instance \cite{Rodriguez-Gomez:2016cem}, it is very 
convenient to obtain the $\N=2$ correlators from the 
$\N=4$ ones by simply expanding (\ref{3.2}) using 
\be
\la{3.7}
\log H(x) = -(1+\gamma_\text{E})\,x^2-\sum_{\ell=2}^\infty (-1)^\ell\frac{\zeta(2\,\ell-1)}{\ell}\,x^{2\,\ell},
\ee
and reconstructing products of higher order traces $\tr a^n$ in the expansion. This procedure is straightforward
and we find $|Z_{1-\text{loop}}|^2 = e^{-S_\text{int}}$ with (see (\ref{3.5}) )
\begin{align}
\la{3.8}
S_\text{int} &= 3\, \zeta (3)\, t_{2,2}-\frac{10}{3}\,\zeta(5)\, 
(3\,t_{2,4}-2\,t_{3,3})+\frac{7}{2}\, \zeta (7)\,
 (4\,t_{2,6}-8\,t_{3,5}+5 \, t_{4,4})\notag \\
 &-\frac{6}{5}\,\zeta (9)\,(15\,t_{2,8}-40\,t_{3,7}+70\,t_{4,6}
 -42 \, t_{5,5})+\dots.
\end{align}
Taking into account relations like (\ref{3.6}) and the rescaling in (\ref{3.4}), this gives immediately the
relation between partition functions
\begin{align}
\la{3.9}
& Z_\sqcd = Z_{\N=4}\,\bigg[1-\frac{3\,(N^4-1)\,\zeta(3)}{16\pi^2\,(\imtau)^2}+
\frac{5\,(N^4-1)(2N^2-1)\,\zeta(5)}{32\pi^3\,N\,(\imtau)^3}\notag \\
&+\bigg(
9\,(N^2+1)(N^2+3)(N^2+5)\,\zeta(3)^2-\frac{35}{N^2}(8N^6+4N^4-3N^2+3)\,\zeta(7)
\bigg)\frac{N^2-1}{512\pi^4\,(\imtau)^4}\notag \\
&+\bigg(
-15\,(N^2+1)(N^2+5)(N^2+7)(2N^2-1)\,\zeta(3)\,\zeta(5)\notag \\
&+\frac{21}{N^2}\,(26N^8+28N^6-3N^4+6N^2-9)\,\zeta(9)
\bigg)\,\frac{N^2-1}{512\pi^5\,N\,(\imtau)^5}+\dots\bigg].
\end{align}

\subsection{Sample 5-loops chiral two-point functions}

Chiral primaries may be computed by taking into account (\ref{3.8}) and the normalization (\ref{3.9}).
Let us give some examples at 5-loop order. This will be useful in the following to explicitly test certain 
differential equations of Toda type. In this section, we consider the following chiral primaries on $\mb R^4$
\be
\la{3.10}
\mc O_2 = \tr\varphi^2, \quad \mc O_{2,2} = (\tr\varphi^2)^2, \quad \mc O_{2,3} = (\tr\varphi^2)\,
\tr\varphi^3, \dots.
\ee
These are related to mixed operators on $S^4$ that we can express conveniently in terms of the matrix
model multiple traces, see (\ref{2.11}),
\begin{align}
\la{3.11}
\mc O_2 &\to t_2 + c_{2;0}(g)\,\mb I,\qquad
\mc O_{2,2} \to t_{2,2}+c_{2,2; 2}(g)\,t_2+c_{2,2; 0}\,\mb I, \notag \\
\mc O_{2,3} &\to t_{2,3}+c_{2,3; 3}(g)\,t_3.
\end{align}
Here $t_{\bm{n}}$ are as in (\ref{3.5}) but we have made explicit the 
rescaling (\ref{3.4}) to make $g$ appear
explicitly in the mixing coefficients. These are determined by Gram-Schmidt orthogonalization. For instance, 
one has in the above
\begin{align}
\la{3.12}
c_{2; 0}(g) &= -\frac{N^2-1}{2}\,\bigg[1-\frac{3 g^4 \,(N^2+1) \zeta (3)}{64 \pi ^4}
+\frac{15 g^6 (N^2+1) (2 \
N^2-1) \zeta (5)}{1024 \pi ^6 N}\notag \\
&+g^8 \bigg(\frac{9 (N^2+1) (N^2+2) \zeta 
(3)^2}{2048 \pi ^8}-\frac{35 (8 N^6+4 N^4-3 N^2+3) \zeta (7)}{16384 \
\pi ^8 N^2}\bigg)\notag \\
&+g^{10}\bigg(\frac{105 (26 N^8+28 N^6-3 N^4+6 N^2-9) \zeta 
(9)}{262144 \pi ^{10} N^3}\notag \\
&-\frac{225 (N^2+1) (N^2+3) (2 N^2-1) \zeta 
(3) \zeta (5)}{65536 \pi ^{10} N}\bigg)+\mc O(g^{12})\bigg],
\end{align}
and similar expansions for the other coefficients, as in 
\begin{align}
\la{3.13}
c_{2,2; 2}(g) &= -(N^2+1)\,\bigg[
1-\frac{3 g^4 (N^2+5) \zeta (3)}{64 \pi ^4}+\frac{15 g^6 (N^2+11) \
(2 N^2-1) \zeta (5)}{1024 \pi ^6 N}\notag \\
&+g^8 \bigg(\frac{9 (N^4+12 N^2+29) \
\zeta (3)^2}{2048 \pi ^8}-\frac{35 (N^2+19) (8 N^6+4 N^4-3 N^2+3) \
\zeta (7)}{16384 \pi ^8 N^2 (N^2+1)}\bigg)\notag \\
&+g^{10} \bigg(\frac{105 (N^2+29) (26 \
N^8+28 N^6-3 N^4+6 N^2-9) \zeta (9)}{262144 \pi ^{10} N^3 \
(N^2+1)}\notag \\
&-\frac{225 (N^2+5) (N^2+15) (2 N^2-1) \zeta (3) \zeta 
(5)}{65536 \pi ^{10} N}\bigg)+\mc O(g^{12})\bigg],\notag \\
c_{2,3; 3}(g) &= -\frac{N^2+5}{2}\,\bigg[
1-\frac{3 g^4 (N^2+7) \zeta (3)}{64 \pi ^4}+\frac{15 g^6 (N^2-1) 
(2 N^4+45 N^2+105) \zeta (5)}{1024 \pi ^6 N (N^2+5)}\notag \\
& +g^8 \bigg
(\frac{9 (N^2+7) (N^2+8) \zeta (3)^2}{2048 \pi ^8}-\frac{35 (8 \
N^8+260 N^6+281 N^4-378 N^2+693) \zeta (7)}{16384 \pi ^8 N^2 \
(N^2+5)}\bigg)\notag \\
& +g^{10}\bigg (\frac{105 (N^2-1) (26 N^8+1180 N^6+3309 \
N^4+3096 N^2+3861) \zeta (9)}{262144 \pi ^{10} N^3 (N^2+5)}\notag \\
& -\frac{225 \
(N^2-1) (N^2+9) (2 N^4+45 N^2+105) \zeta (3) \zeta (5)}{65536 \pi 
^{10} N (N^2+5)}\bigg)+\mc O(g^{12})
\bigg].
\end{align}
After resolution of mixing, correlators can be computed by simply replacing 
 the r.h.s. in (\ref{3.11}) and using the Gaussian correlators, cf. (\ref{3.6}). Since we want to 
 match the normalization in \cite{Gerchkovitz:2016gxx} for the 2-point functions
 of the form $\langle \mc O_{\bm n}(\infty)\,\bar{\mc O}_{\bm n}(0)\rangle$, we also have to 
 multiply the matrix model expression by $2^{|\bm{n}|} = 2^{n_1+n_2+\dots}$ including also 
 the factor $(\frac{g^2}{4\pi})^{|\bm{n}|}$.
 
In particular, let us present the complete five-loop expressions (with generic $N$) 
 of two correlators computed at 3-loops in \cite{Gerchkovitz:2016gxx} for $N=3$ and $N=4$. These are
 \begin{align}
 \la{3.14}
 & \langle \mc O_{2,2}\,\overline{\mc O}_{2,2}\rangle = 8\,(N^4-1)\,\left(\frac{g^2}{4\pi}\right)^4\,\bigg[
 1-\frac{9 g^4 (N^2+3) \zeta (3)}{32 \pi ^4}+\frac{15 g^6 (N^2+6) (2 \
N^2-1) \zeta (5)}{128 \pi ^6 N}\notag \\
&+g^8 \bigg(\frac{9 (29 N^4+204 N^2+355) \
\zeta (3)^2}{4096 \pi ^8}-\frac{175 (N^2+10) (8 N^6+4 N^4-3 N^2+3) \
\zeta (7)}{8192 \pi ^8 N^2 (N^2+1)}\bigg)\notag \\
&+g^{10} \bigg(\frac{315 (N^2+15) (26 \
N^8+28 N^6-3 N^4+6 N^2-9) \zeta (9)}{65536 \pi ^{10} N^3 \
(N^2+1)}\notag \\
&-\frac{135 (2 N^2-1) (7 N^4+78 N^2+211) \zeta (3) \zeta 
(5)}{16384 \pi ^{10} N}\bigg)+\mc O(g^{12})
 \bigg],
 \end{align}
and 
\begin{align}
 \la{3.15}
 & \langle \mc O_{2,3}\,\overline{\mc O}_{2,3}\rangle = 
 6\,\frac{(N^2-1)(N^2-4)(N^2+5)}{N}\,\left(\frac{g^2}{4\pi}\right)^5\,\bigg[ 
 1-\frac{9 g^4 (N^2+5) \zeta (3)}{32 \pi ^4}\notag \\
 &+\frac{5 g^6 (N^2-1) \
(46 N^4+703 N^2+1525) \zeta (5)}{1024 \pi ^6 N (N^2+5)}+g^8 \bigg(\frac{9 \
(7 N^4+75 N^2+215) \zeta (3)^2}{1024 \pi ^8}\notag \\
&-\frac{35 (73 N^8+1501 \
N^6+1537 N^4-2043 N^2+3900) \zeta (7)}{16384 \pi ^8 N^2 \
(N^2+5)}\bigg)\notag \\
&+g^{10}\bigg(\frac{63 (N-1) (N+1) (452 N^8+13312 N^6+36365 \
N^4+34180 N^2+41835) \zeta (9)}{262144 \pi ^{10} N^3 \
(N^2+5)}\notag \\
&-\frac{45 (N-1) (N+1) (78 N^6+1721 N^4+11042 N^2+18575) \zeta 
(3) \zeta (5)}{32768 \pi ^{10} N (N^2+5)}\bigg)+\mc O(g^{12})
  \bigg].
 \end{align}

 Specialization to $SU(3)$ gives
 {\small
 \begin{align}
 \la{3.16}
 & \langle \mc O_{2,2}\,\overline{\mc O}_{2,2}\rangle_{N=3} = \left(\frac{g^2}{4\pi}\right)^4\,\bigg[
 640-\frac{2160 g^4 \zeta (3)}{\pi ^4}+\frac{6375 g^6 \zeta (5)}{\pi 
^6}+g^8 \bigg(\frac{51075 \zeta (3)^2}{8 \pi ^8}-\frac{1699075 \zeta 
(7)}{96 \pi ^8}\bigg)\notag \\
&+g^{10} \bigg(\frac{417375 \zeta (9)}{8 \pi 
^{10}}-\frac{707625 \zeta (3) \zeta (5)}{16 \pi ^{10}}\bigg)+\mc O(g^{12})\bigg], \notag \\ 
 & \langle \mc O_{2,3}\,\overline{\mc O}_{2,3}\rangle_{N=3} = \left(\frac{g^2}{4\pi}\right)^5\,\bigg[
 1120-\frac{4410 g^4 \zeta (3)}{\pi ^4}+\frac{144725 g^6 \zeta (5)}{12 \
\pi ^6}+g^8 \bigg(\frac{458955 \zeta (3)^2}{32 \pi ^8}-\frac{12273275 \
\zeta (7)}{384 \pi ^8}\bigg)\notag \\
&+g^{10} \bigg(\frac{11641175 \zeta (9)}{128 \pi 
^{10}}-\frac{2945775 \zeta (3) \zeta (5)}{32 \pi ^{10}}\bigg)+\mc O(g^{12})\bigg],
 \end{align}
 }
 extending  the 3-loops results in Eq.(3.45) and Eq.(3.56, first line) of  \cite{Gerchkovitz:2016gxx}.
Similarly, specialization to $SU(4)$ gives
{\small
 \begin{align}
 \la{3.17}
 & \langle \mc O_{2,2}\,\overline{\mc O}_{2,2}\rangle_{N=4} = \left(\frac{g^2}{4\pi}\right)^4\,\bigg[
 2040-\frac{43605 g^4 \zeta (3)}{4 \pi ^4}+\frac{1304325 g^6 \zeta 
(5)}{32 \pi ^6}\notag \\
& +g^8 \bigg(\frac{25343685 \zeta (3)^2}{512 \pi 
^8}-\frac{1151616375 \zeta (7)}{8192 \pi ^8}\bigg)+g^{10} \
\bigg(\frac{266283200925 \zeta (9)}{524288 \pi ^{10}}-\frac{3469385925 \
\zeta (3) \zeta (5)}{8192 \pi ^{10}}\bigg)+\mc O(g^{12})\bigg], \notag \\ 
 & \langle \mc O_{2,3}\,\overline{\mc O}_{2,3}\rangle_{N=4} = \left(\frac{g^2}{4\pi}\right)^5\,\bigg[
 5670-\frac{535815 g^4 \zeta (3)}{16 \pi ^4}+\frac{248558625 g^6 \zeta 
(5)}{2048 \pi ^6}\notag \\
& +g^8 \bigg(\frac{81826605 \zeta (3)^2}{512 \pi 
^8}-\frac{13344472575 \zeta (7)}{32768 \pi ^8}\bigg)
+g^{10} \bigg
(\frac{11997966800925 \zeta (9)}{8388608 \pi ^{10}}-\frac{87052714875 \
\zeta (3) \zeta (5)}{65536 \pi ^{10}}\bigg)+\mc O(g^{12})\bigg],
 \end{align}
 }
 extending the 3-loops results in Eq.(3.46) and Eq.(3.57) of  \cite{Gerchkovitz:2016gxx}.

\section{Toda equation as a constraint: the function $F(\lambda; N)$ at $\mc O(\lambda^{10})$}
\la{4}

The explicit results derived in the previous sections are not useful for the study of the large $n$
limit of $g_{2n}$. Indeed, the computational complexity grows quickly with $n$, and an analytic
treatment of the $n$-dependence is mandatory. We now show that this is fully provided by the Toda equation.
The tree level value of $g_{2n}$ is equal to its $\N=4$ limit and reads 
\cite{Rodriguez-Gomez:2016cem}
\be
\la{4.1}
\left. g_{2n}\right|_{\N=4} = \frac{n!\,2^{2n}}{(\imtau)^{2n}}\,
\frac{\Gamma\left(\frac{N^2-1}{2}+n\right)}
{\Gamma\left(\frac{N^2-1}{2}\right)}.
\ee
Explicit results, as in (\ref{2.17}),  suggest that the perturbative corrections takes the following form 
\begin{align}
\la{4.2}
\left. g_{2n}\right|_{\sqcd} &= \frac{n!\,2^{2n}}{(\imtau)^{2n}}\,
\frac{\Gamma\left(\frac{N^2-1}{2}+n\right)}
{\Gamma\left(\frac{N^2-1}{2}\right)}\,\bigg[1+
\mc{A}_2\,n\,\frac{n+\xi_{2,0}}{(\imtau)^2}+
\mc{A}_3\,n\,\frac{n^2+\xi_{3,1}\,n+\xi_{3,0}}{(\imtau)^3}+\notag \\
&+\mc{A}_4\,n\,\frac{n^3+\xi_{4,2}\,n^2+\xi_{4,1}\,n+\xi_{4,0}}{(\imtau)^4}+\dots\bigg],
\end{align}
where the constants $\mc A_k$ and $\xi_{p,q}$ are functions of $N$, but not of $n$ or the coupling.
This  simple structure may be plugged in the Toda equation (\ref{2.6}). Remarkably, the Toda equation is 
strong enough to fix all these constants $\xi_{p,q}$ while the normalization $\mc A_k$ remains free.
For instance, the first two corrections in (\ref{4.2}) take the form 
\begin{align}
\la{4.3}
\left. g_{2n}\right|_{\sqcd} &= \frac{n!\,2^{2n}}{(\imtau)^{2n}}\,
\frac{\Gamma\left(\frac{N^2-1}{2}+n\right)}
{\Gamma\left(\frac{N^2-1}{2}\right)}\,\bigg[1+
\mc{A}_2\,n\,\frac{n+\frac{N^2-1}{2}}{(\imtau)^2}\notag \\
&+\mc{A}_3\,n\,\frac{n^2+\frac{3}{4}\,(N^2-1)\,n+\frac{1}{20}(
3N^4-3N^2+4)}{(\imtau)^3}+\dots\bigg],
\end{align}
The Toda equation may be imposed at high order in the perturbative expansion. 
In general, the coefficients $\xi_{p,q}$ have also a dependence on the constants $\mc A_k$. For instance, at 
the next order, one finds
\begin{align}
\la{4.4}
\xi_{4,2} &= N^2-1+\frac{\mc A_2^2}{\mc A_4}, \notag \\
\xi_{4,1} &= \frac{1}{28}\,\bigg(23-12N^2+9N^4-(20-20N^2+N^4)\,
\frac{\mc A_2^2}{\mc A_4}\bigg), \notag \\
\xi_{4,0} &= \frac{1}{504}\,\bigg(18\,(-8+7N^2+N^6)+(163-147N^2+77N^4-9N^6)\,
\frac{\mc A_2^2}{\mc A_4}
\bigg).
\end{align}
The final expression 
may be evaluated at $n=1$ in terms of the correlator $g_2$ which is related to the double derivative
of the partition function. This allows to fix the constants $\mc A_k$. Explicitly, one finds the following first 
five values \footnote{
Although algorithmic, the procedure requires some careful coding when pushed to high orders. 
The complete solutions at order $\mc O(g^{20})$ is available under request.}
\begin{align}
\la{4.5}
\mc A_2 &= -\frac{9 \zeta (3)}{2 \pi ^2}, \quad 
\mc A_3 = \frac{25 (2 N^2-1) \zeta (5)}{\pi ^3 N (N^2+3)}, \notag \\
\mc A_4 &= \frac{81 \zeta (3)^2}{8 \pi ^4}-\frac{1225 (8 N^6+4 N^4-3 N^2+3) \
\zeta (7)}{16 \pi ^4 N^2 (N^2+1) (N^2+3) (N^2+5)},\notag \\
\mc A_5 &= \frac{1323 (26 N^8+28 N^6-3 N^4+6 N^2-9) \zeta (9)}{4 \pi ^5 N^3 \
(N^2+1) (N^2+3) (N^2+5) (N^2+7)}-\frac{225 (2 N^2-1) \zeta (3) \zeta 
(5)}{2 \pi ^5 N (N^2+3)},\notag \\
\mc A_6 &= \frac{11025 (8 N^6+4 N^4-3 N^2+3) \zeta (3) \zeta (7)}{32 \pi ^6 N^2 \
(N^2+1) (N^2+3) (N^2+5)}\notag \\
&-\frac{17787 (122 N^{10}+280 N^8+48 N^6-15 \
N^4+45) \zeta (11)}{16 \pi ^6 N^4 (N^2+1) (N^2+3) (N^2+5) (N^2+7) \
(N^2+9)}\notag \\
&+\frac{25 (100 N^{12}+2331 N^{10}+13070 N^8+20941 N^6+20985 \
N^4+14138 N^2+7875) \zeta (5)^2}{2 \pi ^6 N^2 (N^2+1) (N^2+3)^2 \
(N^2+5) (N^2+7) (N^2+9)}\notag \\
&-\frac{243 \zeta (3)^3}{16 \pi ^6}.
\end{align}
The full expression of $F(g, n; N)$ is clearly unwieldy. With the definition 
\be
\la{4.6}
F(g, n; N) = 1+\sum_{k=2}^\infty \frac{f_k(n; N)}{(\imtau)^k},
\ee
the $k=2,3$ terms have been written in   (\ref{2.17}). The next two are 
\begin{align}
\la{4.7}
f_4(n; N) &= \frac{9\,n\,(13-45n+36n^2+36n^3+12(-1+3n+3n^2)N^2+(11+9n)N^4)\,\zeta(3)^2}{32\pi^4}
\notag \\
& -\frac{175\,n\,
(8-7n+14n^2+(1+7n)N^2+N^4)(3-3N^2+4N^4+8N^6)\,\zeta(7)}{64\pi^4\,N^2(N^2+1)(N^2+3)(N^2+5)},\notag\\
f_5(n; N) &= -\frac{45n(-1+2n+N^2)(-1+2N^2)}{32\pi^5 N(N^2+3)}[57-52n+90n^2+40n^3\notag \\
& +6(1+9n+5n^2)N^2+3(3+2n)N^4]\,\zeta(3)\,\zeta(5)\notag \\
&\frac{21n}{128\pi^5 N^3(N^2+1)(N^2+3)(N^2+5)(N^2+7)}\,[\notag \\
&7 (67 - 330 n + 560 n^2 - 360 n^3 + 288 n^4) + 
 30 (-17 + 77 n - 56 n^2 + 84 n^3) N^2 \notag \\
 &+ 
 10 (38 - 21 n + 112 n^2) N^4 + 
 30 (1 + 7 n) N^6 + 15 N^8
]\,\zeta(9).
\end{align}
Additional terms are increasingly more involved, but with roughly the same structure. At this level, we did not
identify any simple regularity. 

\subsection{Large R-charge limit}

From our results for the functions $f_k$ in (\ref{4.6}) , 
we can take the limit (\ref{1.2}).  It is convenient to present the result in the following 
logarithmic form 
\footnote{Here $\bm{s} = (s_1, s_2, \dots)$ is a multi-index with non-negative integer values.
We shall also adopt the notation $s_k^p$ to denote repetitions of $p$ instances of $s_k$. So, for instance, 
$\zeta(3^2, 5^3) = \zeta(3,3,5,5,5) = \zeta(3)^2\zeta(5)^3$ and so on.} in order to emphasize possible
exponentiation properties
\begin{align}
\la{4.8}
\log F(\lambda; N) &= \sum_{\ell=2}^\infty \left(\frac{\lambda}{8\pi^2}\right)^\ell\,\sum_{\bm{s}} 
F^{(\ell)}_{\bm s}(N)\,\zeta(\bm{s}),
\qquad \zeta(\bm{s}) = \zeta(s_1)\,\zeta(s_2)\,\dots.
\end{align}
At order $\mc O(\lambda^{10})$, the functions $F^{(\ell)}_{\bm s}(N)$ are
\begin{align}
\la{4.9}
F^{(2)}_{3}(N) &= -18,\notag \\
F^{(3)}_{5}(N) &= \frac{200 (2 N^2-1)}{N (N^2+3)},\notag \\
F^{(4)}_{7}(N) &= -\frac{1225 (8 N^6+4 N^4-3 N^2+3)}{N^2 (N^2+1) (N^2+3) (N^2+5)}, \notag \\
F^{(5)}_{9}(N) &= \frac{10584 (26 N^8+28 N^6-3 N^4+6 N^2-9)}{N^3 (N^2+1) (N^2+3) \
(N^2+5) (N^2+7)}, \notag \\
F^{(6)}_{5^2}(N) &= \frac{184800 (N^2-4) (N^6-N^4-43 N^2-37)}{(N^2+1) (N^2+3)^2 \
(N^2+5) (N^2+7) (N^2+9)},\notag \\
F^{(6)}_{11}(N) &= -\frac{71148 (122 N^{10}+280 N^8+48 N^6-15 N^4+45)}{N^4 (N^2+1) \
(N^2+3) (N^2+5) (N^2+7) (N^2+9)}. \notag \\
F^{(7)}_{5,7}(N) &= -\frac{960960 (N^2-4) (22 N^8+11 N^6-1167 N^4-531 N^2+705)}{N \
(N^2+1) (N^2+3)^2 (N^2+5) (N^2+7) (N^2+9) (N^2+11)},\notag \\
F^{(7)}_{13}(N) &=\frac{8833968 (34 N^{10}+110 N^8-29 N^6+20 N^4-15)}{N^5 (N^2+3) \
(N^2+5) (N^2+7) (N^2+9) (N^2+11)},
\end{align}
with the $\ell=8,9,10$ functions being written in App.~(\ref{app:F}). One can check that 
all terms associated with multiple products of zeta functions vanish for $N=2$. For this value, 
our expression reduces to 
\begin{align}
\la{4.10}
\log F(\lambda; 2) &= -\frac{9 \lambda ^2 \zeta (3)}{32 \pi ^4}+\frac{25 \lambda ^3 \zeta 
(5)}{128 \pi ^6}-\frac{2205 \lambda ^4 \zeta (7)}{16384 \pi 
^8}+\frac{3213 \lambda ^5 \zeta (9)}{32768 \pi ^{10}}\notag \\
& -\frac{78771 \
\lambda ^6 \zeta (11)}{1048576 \pi ^{12}}+\frac{250965 \lambda ^7 \
\zeta (13)}{4194304 \pi ^{14}}-\frac{105424605 \lambda ^8 \zeta 
(15)}{2147483648 \pi ^{16}}\notag \\
& +\frac{265525975 \lambda ^9 \zeta 
(17)}{6442450944 \pi ^{18}}-\frac{12108123027 \lambda ^{10} \zeta 
(19)}{343597383680 \pi ^{20}}+\mc O(\lambda ^{11}),
\end{align}
Extending the $\mc O(\lambda^5)$ results of \cite{Bourget:2018obm}, see (\ref{2.19}),
 and confirming their conjecture 
for the exponentiation in terms of simple $\zeta$-functions, at least at this order. Other instances have a more
complicated structure with products of $\zeta$-functions. For instance, in $SU(3)$ one has 
\begin{align}
\la{4.11}
\log F(\lambda; 3) &=-\frac{9 \lambda ^2 \zeta (3)}{32 \pi ^4}+\frac{425 \lambda ^3 \zeta 
(5)}{2304 \pi ^6}-\frac{17885 \lambda ^4 \zeta (7)}{147456 \pi 
^8}+\frac{5565 \lambda ^5 \zeta (9)}{65536 \pi ^{10}}\notag \\
& +\lambda ^6 \
\bigg(\frac{1925 \zeta (5)^2}{14155776 \pi ^{12}}-\frac{2668897 \zeta 
(11)}{42467328 \pi ^{12}}\bigg)+\lambda ^7 \bigg(\frac{32984237 \zeta 
(13)}{679477248 \pi ^{14}}-\frac{5005 \zeta (5) \zeta (7)}{14155776 \
\pi ^{14}}\bigg)\notag \\
& +\lambda ^8 \bigg(\frac{35035 \zeta (7)^2}{150994944 \pi 
^{16}}+\frac{146575 \zeta (5) \zeta (9)}{402653184 \pi 
^{16}}-\frac{2245755655 \zeta (15)}{57982058496 \pi ^{16}}\bigg)\notag\\
&+\lambda^9 \bigg(-\frac{1519375 \zeta (5)^3}{1174136684544 \pi 
^{18}}-\frac{3488485 \zeta (7) \zeta (9)}{7247757312 \pi 
^{18}}-\frac{546184925 \zeta (5) \zeta (11)}{1565515579392 \pi 
^{18}}\notag \\
&+\frac{669686057755 \zeta (17)}{21134460321792 \pi 
^{18}}\bigg)+\lambda ^{10}\bigg(\frac{8083075 \zeta (5)^2 \zeta 
(7)}{1565515579392 \pi ^{20}}+\frac{77643709 \zeta 
(9)^2}{309237645312 \pi ^{20}}\notag \\
& +\frac{2905703801 \zeta (7) \zeta 
(11)}{6262062317568 \pi ^{20}}+\frac{4074100745 \zeta (5) \zeta 
(13)}{12524124635136 \pi ^{20}}-\frac{29805018472801 \zeta 
(19)}{1127171217162240 \pi ^{20}}\bigg)+\mc O(\lambda ^{11}).
\end{align}

\section{Decoupled Toda equation for the $\tr\varphi^3$ tower}
\la{5}

Given the effectiveness of  Toda equation in the high order calculation of $g_{2n}$, it is natural
to ask whether  similar cases may be treated with the same approach.
As we explained, the Toda equation is a consequence of the determinant representation (\ref{2.12}).
On the other hand, this appears to be related to the one-dimensional mixing structure of $(\tr\varphi^2)^n$
in the $\N=4$ limit. From this point of view, a special case is that of $g_{2n}^\Phi$, see (\ref{2.8}), 
where $\Phi=\tr\varphi^3$. To  simplify notation, we shall denote this 2-point function by the special notation
$\widehat{g}_{2n}$.

The important (easy)  remark is that $(\tr\varphi^2)^n\tr\varphi^3$ can only mix with similar
operators with $n'<n$ 
in the $\N=4$ theory. This is clear from the normal-ordering interpretation of mixing 
discussed in \cite{Billo:2017glv}
and relations like (pairing stands for Wick contraction)
\be
\la{5.1}
\wick{1}{{\rm{Tr}} <1 \varphi^2 {\rm{Tr}} >1 \varphi^3}\sim \tr\varphi^3, \qquad
\wick{1}{{\rm{Tr}} ( <1 \varphi >1 \varphi \varphi)}\sim \tr\varphi = 0, \ \text{etc.}
\ee
This simple mixing pattern 
suggests
the validity of the Toda equation (\ref{2.9}), {\em i.e.}
\be
\la{5.2}
\partial_\tau\partial_{\bar\tau}\log \widehat{g}_{2n} = 
\frac{\widehat{g}_{2n+2}}{\widehat{g}_{2n}}
-\frac{\widehat{g}_{2n}}{\widehat{g}_{2n-2}}-g_2.
\ee
Before showing how to use (\ref{5.2})  to constrain
the R-charge dependence, let us begin with some explicit check of (\ref{5.2}) at low values of $n$.
The correlator $\widehat{g}_0 = \langle \tr\varphi^3\,\tr\bar\varphi^3\rangle$ is the easiest since
there is no mixing to account for. One finds, at order $\mc O(g^{10})$, 
\begin{align}
\la{5.3}
\widehat{g}_0 &= \frac{3 (N^2-4) (N^2-1)}{N}\,\left(\frac{g^2}{4\pi}\right)^3\,\bigg[
1-\frac{9 g^4 (N^2+3) \zeta (3)}{64 \pi ^4}+\frac{5 g^6 (N^2-1) \
(22 N^2+53) \zeta (5)}{1024 \pi ^6 N}\notag \\
& +g^8 \bigg(\frac{27 (3 N^4+20 N^2+37) \
\zeta (3)^2}{4096 \pi ^8}-\frac{105 (11 N^6+12 N^4-16 N^2+29) \zeta 
(7)}{16384 \pi ^8 N^2}\bigg)\notag \\
&+g^{10} \bigg(\frac{63 (N^2-1) (192 N^6+552 \
N^4+515 N^2+645) \zeta (9)}{262144 \pi ^{10} N^3}\notag \\
& -\frac{45 (N^2-1)  (50 N^4+373 N^2+633) \zeta (3) \zeta (5)}{65536 \pi ^{10} \
N}\bigg)+\mc O(g^{12})
\bigg].
\end{align}
The next correlator is 
\be
\la{5.4}
\widehat{g}_2 = \langle\tr\varphi^2\,\tr\varphi^3\,\tr\bar\varphi^2\,\tr\bar\varphi^3\rangle,
\ee
and has been computed in (\ref{3.15}). 
Working out the mixing coefficients appearing in the correlators for $n=2,3$ one finds the explicit
results collected in App.~(\ref{app:hat}). Plugging these results in (\ref{5.2}) and taking $g_2$ from 
(\ref{4.1})  and (\ref{4.6}), one checks that the Toda equation (\ref{5.2}) for $\widehat{g}_{2n}$
is indeed satisfied at this order. 

The case of more complicated towers has roused some debate in the past. A detailed account
of what happens to decoupling is discussed and clarified in  a simple example in App.~(\ref{app:dec}).

\subsection{Using the Toda equation as a constraint for the $\widehat{g}_{2n}$} 

We can propose an Ansatz for the perturbative corrections to $\widehat{g}_{2n}$ correlators
similar to (\ref{4.2}).
It reads
\begin{align}
\la{5.5}
\widehat{g}_{2n} &=  \frac{c_N\,n!\,2^{2n}}{(\imtau)^{2n+3}}\,
\frac{\Gamma\left(\frac{N^2-1}{2}+3+n\right)}
{\Gamma\left(\frac{N^2-1}{2}+3\right)}\,\bigg[1+
\mc{B}_2\,\frac{n^2+\eta_{2,1}\,n+\eta_{2,0}}{(\imtau)^2}+
\mc{B}_3\,\frac{n^3+\eta_{3,2}\,n^2+\eta_{3,1}\,n+\eta_{3,0}}{(\imtau)^3}+\notag \\
&+\mc{B}_4\,\frac{n^4+\eta_{4,3}\,n^3+\eta_{4,2}\,n^2+\eta_{4,1}\,n+\eta_{4,0}}{(\imtau)^4}+\dots\bigg],
\end{align}
where $c_N = \frac{3\,(N^2-4)(N^2-1)}{N}$. Compared with (\ref{4.2}), we see that in this case
the corrections do not vanish for $n=0$ because 
$\widehat{g}_0 = \langle \tr\varphi^3\,\tr\bar\varphi^3\rangle$
is clearly non-trivial.
Again, imposing the Toda equation, we can fix all 
the $\eta$-coefficients but not the $\mc B_k$ ones. For instance, one finds at first order
\be
\la{5.6}
\eta_{2,1} = \frac{1}{2}(N^2+5),\qquad \eta_{2,0} = \frac{1}{48}\,\bigg[
2\,(35+12N^2+N^4)+\frac{9(N^4-1)\zeta(3)}{\pi^2\,\mc B_2}\bigg].
\ee
Matching the Ansatz (\ref{5.5}) to the $n=0$ case gives then 
\be
\la{5.7}
\mc B_2 = -\frac{9\,\zeta(3)}{2\pi^2}.
\ee
We carried on this procedure up to $\mc O(g^{14})$. The ratio SQCD/$\N=4$ is now
expressed by 
\be
\la{5.8}
\widehat{F}(g, n; N) = \frac{\left. \widehat{g}_{2n} \right|_\text{SQCD}}
{\left. \widehat{g}_{2n} \right|_{\N=4}} = 1+\sum_{k=2}^\infty \frac{\widehat{f}_k(n; N)}{(\imtau)^k},
\ee
where
\begin{align}
\la{5.9}
f_2(n; N) &= -\frac{9 (n+1)(N^2+2 n+3) \zeta (3) }{4 \pi ^2},\notag \\
f_3(n; N) &= \frac{5(N^2-1)}{16\pi^3 N(N^2+5)(N^2+7)(N^2+9)}\,[
105 (3 + 2 n) (53 + 90 n + 40 n^2) \notag \\
&+ (14509 + 30780 n + 19800 n^2 + 
    3600 n^3) N^2 + (4259 + 7952 n + 3300 n^2 + 
    160 n^3) N^4 \notag \\
    &+ (515 + 804 n + 
    120 n^2) N^6 + 2 (11 + 12 n) N^8
]\,\zeta(5), \notag \\
f_4(n; N) &= \frac{9}{32\pi^4}[
(3 + 2 n) (74 + 153 n + 99 n^2 + 18 n^3) + 
 12 (1 + n) (10 + 12 n + 3 n^2) N^2 \notag \\
 & + (18 + 29 n + 9 n^2) N^4]\zeta(3)^2 -\frac{35}{64\pi^4 N^2(N^2+5)(N^2+7)(N^2+9)(N^2+11)}
 \notag \\
 & [3465 (1 + n) (3 + 2 n) (29 + 35 n + 14 n^2) \notag \\
 & -  3 (688 + 24675 n + 62370 n^2 + 55860 n^3 + 
    17640 n^4) N^2 \notag \\
    & + 
 2 (33327 + 98135 n + 118060 n^2 + 71890 n^3 + 
    19670 n^4) N^4 \notag \\
    & + 
 5 (33429 + 91204 n + 95074 n^2 + 44268 n^3 + 
    7280 n^4) N^6 \notag \\
     & + (74319 + 182900 n + 148445 n^2 + 
    42000 n^3 + 1120 n^4) N^8 \notag \\
    & + (13446 + 29405 n + 
    15540 n^2 + 1120 n^3) N^{10} \notag \\
     & + 
 4 (273 + 505 n + 90 n^2) N^{12} + (33 + 
    40 n) N^{14}]\,\zeta(7).
\end{align}
and so on. \footnote{Again, the next functions are available under request.}

\subsection{Large R-charge limit}

In the large R-charge limit  (\ref{1.2}) we have the following expansion of the logarithm of 
the function $\widehat F(\lambda, N) = \lim_{n\to \infty} \widehat{F}(\sqrt{\lambda/n}, n; N)$
{\small
\begin{align}
\la{5.10}
& \log  \widehat{F}(\lambda; N) = 
-\frac{9 \lambda ^2 \bm{\zeta (3)}}{32 \pi ^4}+\frac{25 \lambda ^3 (N^2-1) \
 (2 N^4+45 N^2+105) \bm{\zeta (5)}}{64 \pi ^6 N (N^2+5) (N^2+7) \
(N^2+9)}\notag \\
& -\frac{1225 \lambda ^4 (8 N^8+260 N^6+281 N^4-378 N^2+693) \
\bm{\zeta (7)}}{4096 \pi ^8 N^2 (N^2+5) (N^2+7) (N^2+9) \
(N^2+11)}\notag \\
&+\frac{1323 \lambda ^5 (N^2-1) (26 N^8+1180 N^6+3309 \
N^4+3096 N^2+3861) \bm{\zeta (9)}}{4096 \pi ^{10} N^3 (N^2+5) (N^2+7) \
(N^2+9) (N^2+11) (N^2+13)}\notag \\
&+\lambda ^6 \bigg(\frac{5775}{8192 \pi ^{12} (N^2+5)^2 (N^2+7)^2 (N^2+9)^2 (N^2+11) \
(N^2+13) (N^2+15)} \notag \\
& (N^{14}+88 N^{12}
 +15 N^{10}-18088 N^8-39661 N^6+1053540 N^4+4281405 N^2+4399500) \
\bm{\zeta (5)^2}\notag \\
& -\frac{17787 (122 N^{12}+6950 N^{10}+24848 N^8+8085 \
N^6-12645 N^4+15345 N^2+32175) \bm{\zeta (11)}}{65536 \pi ^{12} N^4 \
(N^2+5) (N^2+7) (N^2+9) (N^2+11) (N^2+13) (N^2+15)}\bigg)\notag \\
& +\lambda ^7 \
\bigg(\frac{552123 (N^2-1)}{131072 \pi ^{14} N^5 \
(N^2+5) (N^2+7) (N^2+9) (N^2+11) (N^2+13) (N^2+15) \
(N^2+17)}\notag \\
& (34 N^{12} +2424 N^{10}+17285 N^8+29655 \
N^6+24450 N^4+41145 N^2+16575) \bm{\zeta (13)}\notag \\
& -\frac{15015 (N^2-1) }{32768 \pi ^{14} N (N^2+5)^2 (N^2+7)^2 (N^2+9)^2 \
(N^2+11) (N^2+13) (N^2+15) (N^2+17)}\notag \\
& (22 N^{14} +2323 N^{12}+6951 \
N^{10}-473938 N^8-1641088 N^6+30589515 N^4\notag \\
 & +136118115 N^2+144364500) \
\bm{\zeta (5) \zeta (7)}\bigg)+\mc O(\lambda ^8),
\end{align}}
where we have emphasized the $\zeta$-functions. 
Compared to $F(\lambda; N)$, the $\zeta(3)$-dependent terms again exponentiate and are independent
on $N$. Apart from this, there are no further special simplifications for low values of $N$.  
The first non-trivial specialized cases  are  $N=3, 4$ that give
\footnote{Here it does not make sense to set $N=2$
since the correlators $\widehat{g}_{2n}$ are zero due to $\tr\varphi^3=0$ in $SU(2)$. This is consistent with 
$c_2=0$ in (\ref{5.5}).}
{\small
\begin{align}
\la{5.11}
& \log\widehat{F}(\lambda; 3) = -\frac{9 \lambda ^2 \zeta (3)}{32 \pi ^4}+\frac{25 \lambda ^3 \zeta 
(5)}{144 \pi ^6}-\frac{15925 \lambda ^4 \zeta (7)}{147456 \pi 
^8}+\frac{147 \lambda ^5 \zeta (9)}{2048 \pi ^{10}}\notag \\
&+\lambda ^6 \
\bigg(\frac{1925 \zeta (5)^2}{14155776 \pi ^{12}}-\frac{8599591 \zeta 
(11)}{169869312 \pi ^{12}}\bigg)+\lambda ^7 \bigg(\frac{3177031 \zeta 
(13)}{84934656 \pi ^{14}}-\frac{5005 \zeta (5) \zeta (7)}{14155776 \
\pi ^{14}}\bigg)+\mc O(\lambda^8),\notag \\
& \log \widehat{F}(\lambda; 4) = -\frac{9 \lambda ^2 \zeta (3)}{32 \pi ^4}+\frac{955 \lambda ^3 \zeta 
(5)}{5888 \pi ^6}-\frac{429289 \lambda ^4 \zeta (7)}{4521984 \pi ^8}+\
\frac{78057 \lambda ^5 \zeta (9)}{1310720 \pi ^{10}}\notag \\
&+\lambda ^6 \bigg
(\frac{146531 \zeta (5)^2}{973969408 \pi ^{12}}-\frac{68971014343 \
\zeta (11)}{1734512476160 \pi ^{12}}\bigg)+\lambda ^7 \bigg(\frac{387146868537 \
\zeta (13)}{13876099809280 \pi ^{14}}-\frac{28131103 \zeta (5) \zeta 
(7)}{77917552640 \pi ^{14}}\bigg)+\mc O(\lambda ^8).
\end{align}
}

\section{Relation with large $N$ factorization}
\la{sec:largeN}

Given our results that hold for all values of the R-charge $n$ and gauge rank $N$, it is interesting to
consider their large $N$ expansion, see also \cite{Baggio:2016skg,Rodriguez-Gomez:2016ijh,Pini:2017ouj}.
To this aim, let us define in this section the conventional 't Hooft coupling 
$\lambda = N\,g^2$, not to be confused with the large R-charge coupling in (\ref{1.2}). We are interested in
\be
\la{6.1}
\mc F_n(\lambda) = \lim_{N\to \infty} F(\sqrt{\lambda/N}, n; N),
\ee
where $F(g, n; N)$ is in (\ref{2.15}). \footnote{
It is important to remark that absence of instanton corrections at large $N$ is based on the assumption
that the instanton moduli integration does not spoil the instanton action exponential suppression as $g\to 0$.
In general, this may be a non-trivial issues, and instantons may lead to large $N$ phase transitions,
see for instance \cite{Gross:1994mr}.
A test of this assumption in $\N=2$ SQCD is discussed in \cite{Passerini:2011fe} where it is shown that 
the one-instanton corrections to the partition function leads to a $\sqrt{N}$ enhancement of the 
exponentially small instanton action factor. Thus, it is negligible at large $N$.}
Using  the results in (\ref{4.6}) , we find 
\be
\la{6.2}
\log \mc F_n(\lambda) = n\,\log\mc F_1(\lambda) ,
\ee
with (we write only the first 7 orders, but we checked (\ref{6.2})  at order $\mc O(\lambda^{10})$)
\begin{align}
\la{6.3}
\log & \mc F_1(\lambda) = -\frac{9 \lambda ^2 \zeta (3)}{64 \pi ^4}+\frac{15 \lambda ^3 \zeta 
(5)}{128 \pi ^6}+\lambda ^4 \bigg(\frac{99 \zeta (3)^2}{8192 \pi 
^8}-\frac{175 \zeta (7)}{2048 \pi ^8}\bigg)+\lambda ^5 \bigg(\frac{4095 \zeta 
(9)}{65536 \pi ^{10}}-\frac{405 \zeta (3) \zeta (5)}{16384 \pi 
^{10}}\bigg)\notag \\
&+\lambda ^6 \bigg(-\frac{189 \zeta (3)^3}{131072 \pi 
^{12}}+\frac{6375 \zeta (5)^2}{524288 \pi ^{12}}+\frac{2835 \zeta (3) 
\zeta (7)}{131072 \pi ^{12}}-\frac{98637 \zeta (11)}{2097152 \pi 
^{12}}\bigg)\notag \\
& +\lambda ^7 \bigg(\frac{4995 \zeta (3)^2 \zeta (5)}{1048576 \pi 
^{14}}-\frac{21595 \zeta (5) \zeta (7)}{1048576 \pi 
^{14}}-\frac{77805 \zeta (3) \zeta (9)}{4194304 \pi 
^{14}}+\frac{153153 \zeta (13)}{4194304 \pi ^{14}}\bigg)\notag \\
&+\lambda ^8 \bigg
(\frac{13203 \zeta (3)^4}{67108864 \pi ^{16}}-\frac{169425 \zeta (3) \
\zeta (5)^2}{33554432 \pi ^{16}}-\frac{38115 \zeta (3)^2 \zeta 
(7)}{8388608 \pi ^{16}}+\frac{568645 \zeta (7)^2}{67108864 \pi 
^{16}}\notag \\
& +\frac{144585 \zeta (5) \zeta (9)}{8388608 \pi 
^{16}}+\frac{2155923 \zeta (3) \zeta (11)}{134217728 \pi 
^{16}}-\frac{7818525 \zeta (15)}{268435456 \pi ^{16}}\bigg)+\dots.
\end{align}
Relation (\ref{6.2}) expresses factorization 
\be
\langle(\tr\varphi^2)^n\,(\tr\overline\varphi^2)^n\rangle\sim \langle\tr\varphi^2
\,\tr\overline\varphi^2\rangle^n,
\ee 
at leading large $N$. 
The finite $N$ corrections to (\ref{6.2})  are $\mc O\left(\frac{n}{N^2},\frac{n^2}{N^2}\right)$,
where we explicitly denoted that the dependence on $n$ is the sum of linear and quadratic pieces. 

A similar analysis for the function $\widehat{\mc F}$
which is the large $N$ limit of the function $\widehat F$ in (\ref{5.8}) shows that (\ref{6.2}) reads now
\be
\la{6.4}
\log \frac{\mc{\widehat F}_n(\lambda)}{\mc{\widehat F}_0(\lambda)} = n\,\log\mc F_1(\lambda),
\ee
where $\log\mc F_1(\lambda)$ is the expression in (\ref{6.3}).
%
Again, this implies the following large $N$ factorization 
\be
\langle(\tr\varphi^2)^n\,\tr\varphi^3\,(\tr\overline\varphi^2)^n\,\tr\overline\varphi^3\rangle
\sim \langle\tr\varphi^2\,\tr\overline\varphi^2\rangle^n\,\langle 
\tr\varphi^3\,\tr\overline\varphi^3\rangle.
\ee
Subleading corrections to large $N$ may of course be extracted easily from the 
general expressions valid for generic $N$.

\section*{Acknowledgement}
\addcontentsline{toc}{section}{Acknowledgement}

We thank A. Bourget, D. Rodriguez-Gomez, and Jorge G. Russo for discussions and 
 kind clarifications  about their work. We also thank A. A. Tseytlin for useful suggestions and 
 comments. 

\appendix

\section{Functions $F^{(\ell)}(N)$ for $\ell=8,9,10$}
\la{app:F}

Let us define 
\be
\text{P}_k = \left(\frac{N^2+1}{2}\right)_k,\qquad (x)_k = \frac{\Gamma(x+k)}{\Gamma(x)}.
\ee
The functions $F^{(\ell)}(N)$ defined in Sec.~(\ref{4}) for $\ell=8,9,10$ are 
given by the following expressions.
\begin{align}
F^{(8)}_{7^2}(N) &= \frac{1576575 (N^2-4) }
{256 N^2 \text{P}_3 \text{P}_7}
(98 N^{14}+928 N^{12}-4151 N^{10}-44359 \
N^8-42036 N^6\notag \\
&+26754 N^4+14553 N^2-16299),\notag \\
F^{(8)}_{5,9}(N) &= \frac{675675 (N^2-4)}{2 N^2 (N^2+3) \text{P}_7}
 (23 N^{10}+70 N^8-1455 N^6-1335 N^4+192 \
N^2-855)\notag \\
F^{(8)}_{15}(N) &= -\frac{41409225 }{256 N^6 \text{P}_7}
(540 N^{14}+3780 N^{12}+4676 N^{10}+440 N^8+329 \
N^6\notag \\
& -735 N^4+735 N^2+315),\notag \\
F^{(9)}_{5^3}(N) &= \frac{3038750 (N^2-4) }{3 N (N^2+3)^2 \text{P}_8}
(2 N^{12}-28 N^{10}-511 N^8+2187 N^6+24479 \
N^4+8701 N^2-7950),\notag \\
F^{(9)}_{7,9}(N) &= -\frac{1276275 (N^2-4) }{32 N^3 \text{P}_3\text{P}_8}
(724 N^{16}+9884 N^{14}-19378 N^{12}-455219 \
N^{10}\notag \\
&-693063 N^8-78342 N^6+24156 N^4-48195 N^2+162729),\notag \\
F^{(9)}_{5,11}(N) &= 
-\frac{10027875 (N^2-4) }{8 N^3 (N^2+3) \text{P}_8}(142 N^{12}+926 N^{10}-9807 N^8
\notag \\
& -23606 N^6-3754 N^4+2184 N^2+7035),
\notag \\
F^{(9)}_{17}(N) &= \frac{29548805 }{32 N^7 
\text{P}_8}(1866 N^{16}+19976 N^{14}+45374 N^{12}\notag \\
& +17804 N^{10}+345 N^8-1260 N^6+4410 N^4-6300 N^2-1575),\notag \\
F^{(10)}_{5^2, 7}(N) &= -\frac{121246125 (N^2-4) (N^2-1)}
{32 N^2 (N^2+3) \text{P}_3 
\text{P}_9}
 (7 N^{16}-16 N^{14}-2593 \
N^{12}-11454 N^{10}\notag \\
&+132421 N^8+888024 N^6+1182001 N^4+282422 \
N^2-148380),\notag \\
F^{(10)}_{9^2}(N) &= \frac{2909907 (N^2-4) }{64 N^4 (N^2+7) 
\text{P}_3 \text{P}_9}
(7697 N^{20}+205301 N^{18}+1156794 \
N^{16}\notag \\
&-6284951 N^{14}-65595428 N^{12}-119443263 N^{10}-65064156 \
N^8\notag \\
&-37063809 N^6-36848169 N^4-7387038 N^2-18683298),\notag \\
F^{(10)}_{7,11}(N) &= \frac{53348295 (N^2-4) }{256 N^4 \text{P}_3 
\text{P}_9}
(3218 N^{18}+60754 N^{16}+34194 N^{14}-2724467 \
N^{12}\notag \\
&-7509568 N^{10}-3899769 N^8+2216346 N^6+1637547 N^4-330750 \
N^2-1099665),\notag \\
F^{(10)}_{5,13}(N) &=\frac{297226215 (N^2-4) }{8 N^4 (N^2+3) 
\text{P}_9}
(111 N^{14}+1214 N^{12}-6968 N^{10}-42034 \
N^8-20548 N^6\notag \\
& +5225 N^4-9555 N^2-8085), \notag \\
F^{(10)}_{19}(N) &=-\frac{2133423721}{640 N^8 \text{P}_9}
 (10688 N^{18}+165378 N^{16}+616452 N^{14}+541822 \
N^{12}\notag \\
&+49410 N^{10}+8235 N^8-11340 N^6-28350 N^4+85050 \
N^2+14175).
\end{align}

\section{The correlators $\widehat{g}_{2n}$ for $n=2,3$}
\la{app:hat}

The explicit expressions for the correlators $\widehat{g}_{2n}$ for $n=2,3$ are 
{\small
\begin{align}
\widehat{g}_4 &= \frac{24 (N^2-1) (N^2-4) (N^2+5) (N^2+7)}{N}\,\left(\frac{g^2}{4\pi}\right)^7\,\bigg[
1-\frac{27 g^4 (N^2+7) \zeta (3)}{64 \pi ^4}\notag \\
&+\frac{5 g^6 (N^2-1)
 (70 N^6+1973 N^4+16886 N^2+32095) \zeta (5)}{1024 \pi ^6 N \
(N^2+5) (N^2+7)}\notag \\
&+g^8 \bigg(\frac{9 (14 N^4+207 N^2+805) \zeta (3)^2}{1024 \
\pi ^8}\notag \\
& -\frac{35 (113 N^{10}+4312 N^8+45949 N^6+41951 N^4-54186 \
N^2+113925) \zeta (7)}{16384 \pi ^8 N^2 (N^2+5) (N^2+7)}\bigg)\notag \\
& +g^{10} \
\bigg(\frac{63 (N^2-1) (712 N^{10}+37376 N^8+575639 N^6+1477845 \
N^4+1403305 N^2+1644195) \zeta (9)}{262144 \pi ^{10} N^3 (N^2+5) \
(N^2+7)}\notag \\
&-\frac{45 (N^2-1) (310 N^8+11471 N^6+154469 N^4+833209 \
N^2+1319325) \zeta (3) \zeta (5)}{65536 \pi ^{10} N (N^2+5) \
(N^2+7)}\bigg)+\mc O(g^{12})
\bigg], \notag \\
\widehat{g}_6 &=\frac{144 (N^2-1) (N^2-4) (N^2+5) (N^2+7) (N^2+9)}{N}\,
\left(\frac{g^2}{4\pi}\right)^9\,\bigg[
1-\frac{9 g^4 (N^2+9) \zeta (3)}{16 \pi ^4}\notag \\
&+\frac{5 g^6 (N^2-1) \
(94 N^8+4007 N^6+62135 N^4+382249 N^2+645435) \zeta (5)}{1024 \pi ^6 \
N (N^2+5) (N^2+7) (N^2+9)}\notag \\
&+g^8 \bigg(\frac{27 (31 N^4+584 N^2+2865) \zeta 
(3)^2}{4096 \pi ^8}\notag \\
& -\frac{105 (51 N^{12}+2903 N^{10}+58654 N^8+416054 \
N^6+336111 N^4-414173 N^2+982800) \zeta (7)}{16384 \pi ^8 N^2 (N^2+5) \
(N^2+7) (N^2+9)}\bigg)\notag \\
& +g^{10}\bigg (\frac{63 (N^2-1) }{262144 \pi ^{10} N^3 (N^2+5) (N^2+7) (N^2+9)}
(972 N^{12}+74304 \
N^{10}+2058613 N^8+21049036 N^6\notag \\
& +50557990 N^4+48478260 N^2+53986905) \
\zeta (9)\notag \\
& -\frac{45 \
(N^2-1) (128 N^{10}+6850 N^8+145489 N^6+1476329 N^4+6789279 \
N^2+9999045) \zeta (3) \zeta (5)}{16384 \pi ^{10} N (N^2+5) (N^2+7) \
(N^2+9)}\bigg)+\mc O(g^{12})
\bigg].
\end{align}
}

\section{Decoupling violations and Toda equation for general towers}
\la{app:dec}

Let us clarify to what extent we may have decoupling and reduction to a single semi-infinite Toda equation
for more general towers of the form 
$\Phi^{(n)} = \mc (\tr\varphi^2)^n\,\Phi$ with $\dim\Phi>3$. Although general discussions
already appeared in \cite{Baggio:2015vxa,Gerchkovitz:2016gxx}, we consider here  in full details the example of 
$\Phi=\tr\varphi^4$ and consider in somewhat more details what is the fate of the 
would-be decoupled
Toda equation. 

Compared with the case $\Phi=\tr\varphi^3$, an important difference 
-- already visible in the $\N=4$ limit -- 
is that Wick contraction
inside $\tr\varphi^4$ brings back to the tower of the identity $\Phi=\mb I$. Let us work at the level of 
matrix model traces $t_{\bm{n}}$, cf. (\ref{3.5}) and  denote by $:T:$, the operator $T$ with
resolved mixing, {\em i.e.} after subtraction of the suitable contraction  ordering contributions,
{\em i.e.} as in the r.h.s. of (\ref{3.11}). \footnote{
Notice that in the notation of \cite{Billo:2018oog}, 
this is denoted $:X:_g$ to emphasize the difference with respect to the
$\mc N=4$ limit.}
At dimension $4+2\,n$, we can consider the two operators 
\be
\mc O_n = :t_{\underbrace{\scriptstyle 2, \dots, 2}_{n}, 4}:\ ,\qquad 
\mc O_n' = :t_{\underbrace{\scriptstyle 2, \dots, 2}_{n+2}}:\ .
\ee
These are {\em natural} in the sense that they map to the simple flat space operators 
\be
\mc O^{\mb R^4}_n = (\tr\varphi^2)^n\,\tr\varphi^4,\qquad 
\mc O^{'\ \mb R^4}_n = (\tr\varphi^2)^{n+2}. 
\ee
Besides, we can construct the operator
\be
\la{C.3}
\widetilde{\mc O}_n = \mc O_n -\kappa_n(g)\,\mc O_n',
\ee
where $\kappa_n(g)$ is chosen such that $\langle \widetilde{\mc O}_n \overline{\mc O}'_n\rangle=0$.
The second term in (\ref{C.3}) is a further subtraction unrelated to $S^4$ mixing, since the two operators have the
same dimension. Instead, it is a loop corrected orthogonalization similar in spirit to the 
tree level diagonalization discussed in \cite{Baggio:2015vxa}, and shown to be problematic due to three loop
corrections in \cite{Gerchkovitz:2016gxx}.

Now, it should be clearly stated that the 2-point function of $\widetilde{\mc O}_n$ 
do satisfy the decoupled Toda equation (\ref{2.9}), contrary to $\mc O_n$.
So, decoupling is possible, but the point is that it is not achieved in a natural way 
because the coefficients $\kappa_n(g)$ in (\ref{C.3}) 
depend on $n$. This means that, for instance, the large $n$ limit has to take into account 
the extra $n$ dependence in $\kappa_n(g)$.

To see this in details, let us work out the explicit expressions of the above operators at level $n=0$, {\em i.e.} dimension 4. We find 
\begin{align}
\mc O_0 &= t_4-\frac{(2 N^2-3)}{N (N^2+1)}\,t_{2,2}\notag \\
& -\frac{25 (N^2-9) (N^2-4)  \zeta (5)}{256 \pi ^6 (N^2+1)} \bigg[t_{2,2}+\frac{3}{20} \
(N^4-1) -\frac{3}{4} (N^2+1) t_2\bigg]\,g^6+\mc O(g^8),\\
\mc O'_0 &= t_{2,2}+\frac{1}{4} (N^4-1)-(N^2+1) t_2\notag \\
&\frac{3  (N^2+1) (N^2+5) \zeta (3) }{64 \pi ^4}\bigg[
t_2-\frac{(N^2-1) \
 (N^2+4)}{2 (N^2+5)}\bigg]\,g^4\notag \\
&-\frac{15  (N^2+1) (N^2+11) (2 N^2-1) \zeta (5) \
}{1024 \pi ^6 \ N}\bigg[t_2-\frac{(N^2-1) (N^2+9)}{2 (N^2+11)}\bigg]\,g^6+\mc O(g^8).
\end{align}
The condition 
$\langle \widetilde{\mc O}_0 \overline{\mc O}'_0\rangle=
\langle [\mc O_0 -\kappa_0(g)\,\mc O_0']\,\overline{\mc O}'_0\rangle = 0$ gives
\be
\kappa_0(g) = \frac{2 N^2-3}{N (N^2+1)}+\frac{25\  (N^2-9) (N^2-4) \zeta 
(5)}{256 \pi ^6 (N^2+1)}\,g^6+\mc O(g^8).
\ee
Repeating the calculation for $n=1,2,3$ gives a similar result but with a modified $\mc O(g^6)$ term
\be
\kappa_n(g) = \frac{2 N^2-3}{N (N^2+1)}+\frac{25\  (N^2-9) (N^2-4)  \zeta 
(5)}{256 \pi ^6 (N^2+1)}\,\alpha_n\,g^6+\mc O(g^8),
\ee
where
\be
\alpha_1 = \frac{N^2+11}{N^2+3},\quad 
\alpha_2 = \frac{(N^2+9)(N^2+15)}{(N^2+3)(N^2+5)},\quad 
\alpha_3 = \frac{(N^2+9)(N^2+11)(N^2+19)}{(N^2+3)(N^2+5)(N^2+7)}.
\ee
As anticipated, the coefficient $\kappa_n(g)$ depends on $n$ starting at three loops. 

Let us now show that the 2-point function $\widetilde{g}_{2n}$ of the $\widetilde{\mc O}_n$ operators is still captured by the Toda equation. We find (the tree level is known, see for instance \cite{Gerchkovitz:2016gxx})
\be
\la{C.9}
\widetilde{g}_{2n} = \frac{4\,(N^2-1)(N^2-4)(N^2-9)}{N^2+1}\,\frac{n!\,2^{2n}}{(\imtau)^{2n+4}}\,
\frac{\Gamma(\tfrac{N^2-1}{2}+4+n)}{\Gamma(\tfrac{N^2-1}{2}+4)}\, (1+\delta\widetilde{g}_{2n}),
\ee
with 
{\small
\begin{align}
\la{C.10}
\delta\widetilde{g}_0 &= -\frac{3 g^4 (N^2+4) \zeta (3)}{16 \pi ^4}+\frac{5 g^6 (N^2-1) \
(8 N^2+3) \zeta (5)}{256 \pi ^6 N}+\mc O(g^8), \notag \\
\delta\widetilde{g}_2 &= -\frac{3 g^4 (7 N^2+43) \zeta (3)}{64 \pi ^4}+\frac{5 g^6 (N^2-1) 
 (7 N^4+127 N^2+45) \zeta (5)}{128 \pi ^6 N (N^2+7)}+\mc O(g^8),\notag \\
\delta\widetilde{g}_4 &=-\frac{3 g^4 (5 N^2+41) \zeta (3)}{32 \pi ^4}+\frac{5 g^6 (N^2-1) 
(20 N^6+689 N^4+6150 N^2+2121) \zeta (5)}{256 \pi ^6 N (N^2+7) \
(N^2+9)}+\mc O(g^8)\notag \\
\delta\widetilde{g}_6 &= -\frac{3 g^4 (13 N^2+133) \zeta (3)}{64 \pi ^4}\notag \\
& +\frac{5 g^6 (N^2-1)
 (13 N^8+672 N^6+11876 N^4+71268 N^2+24051) \zeta (5)}{128 \pi \
^6 N (N^2+7) (N^2+9) (N^2+11)}+\mc O(g^8).
\end{align}
}
Using (\ref{C.9})  and (\ref{C.10}) , one can check  that the first instances of the Toda equation (\ref{2.9}) 
are satisfied. This includes the effect of the 3-loop $\zeta(5)$ terms which are responsible for the $g$-dependent
mixing discussed in \cite{Gerchkovitz:2016gxx}. This means that it is easy to obtain closed expressions for the 
expansions in (\ref{C.10}) by the methods discussed in the main text. For instance, the first correction
in (\ref{C.10}) is  
\be
\delta\widetilde{g}_{2n} = -\frac{3\,[16+21\,n+6\,n^2+(4+3\,n)\,N^2]}{64\pi^4}
\,\zeta(3)\,g^4+\dots,
\ee
while higher order corrections may be treated as we discussed the $\mb I$ and $\tr\varphi^3$ towers.

\bibliography{BT-Biblio}
\bibliographystyle{JHEP}

\end{document}